**Hydrogen Bond Networks Near Supported Lipid Bilayers from Vibrational Sum**

**Frequency Generation Experiments and Atomistic Simulations**

Merve Doğangün,[a,#] Paul E. Ohno,[a,#] Dongyue Liang,[b] Alicia C. McGeachy,[a] Ariana Gray Be,[a]

Naomi Dalchand,[a] Tianzhe Li,[a] Qiang Cui,[b,c*] and Franz M. Geiger[a,*]

[a]Department of Chemistry, Northwestern University, 2145 Sheridan Road, Evanston, IL 60660;

[b]Department of Chemistry, University of Wisconsin-Madison, Madison, WI, 53706, USA;

[c]Department of Chemistry, Boston University, 590 Commonwealth Ave., Boston, MA 02215,

USA

*Authors to whom correspondence should be addressed: qiangcui@bu.edu and

geigerf@chem.northwestern.edu

**ABSTRACT.** We report vibrational sum frequency generation (SFG) spectra in which the C–H stretches of lipid alkyl tails in fully hydrogenated single- and dual-component supported lipid bilayers are detected along with the O–H stretching continuum above the bilayer. As the salt concentration is increased from ~10 μM to 0.1 M, the SFG intensities in the O–H stretching region decrease by a factor of 2, consistent with significant absorptive-dispersive mixing between $\chi^{(2)}$ and $\chi^{(3)}$ contributions to the SFG signal generation process from charged interfaces. A method for estimating the surface potential from the second-order spectral lineshapes (in the OH stretching region) is presented and discussed in the context of choosing truly zero-potential reference states. Aided by atomistic simulations, we find that the strength and orientation distribution of the hydrogen bonds over the purely zwitterionic bilayers are largely invariant between sub-micromolar and hundreds of millimolar concentrations. However, specific interactions between water molecules and lipid headgroups are observed upon replacing phosphocholine (PC) lipids with negatively charged phosphoglycerol (PG) lipids, which coincides with SFG signal intensity reductions in the 3100 cm$^{-1}$ to 3200 cm$^{-1}$ frequency region. The atomistic simulations show that this outcome is consistent with a small, albeit statistically significant, decrease in the number of water molecules adjacent to both the lipid phosphate and choline moieties per unit area, supporting the SFG observations. Ultimately, the ability to probe hydrogen-bond networks over lipid bilayers holds the promise of opening paths for understanding, controlling, and predicting specific and non-specific interactions between membranes and ions, small molecules, peptides, polycations, proteins, and coated and uncoated nanomaterials.

**I. Introduction.**   The structure of water over lipid membranes is of interest for a variety of reasons that are rooted in fundamental scientific interest and connect all the way to biological function and technological applications.[1-6] Specific questions pertain to whether there exist populations of interfacial water molecules that can undergo hydrogen-bond (H-bond) interactions with certain membrane constituents that can be strengthened or weakened with variations in ionic strength, or, as indicated by molecular dynamics simulations,[2] whether some population of water molecules exists that may interact specifically with certain lipid headgroups over others.

While interface-specific vibrational spectroscopic approaches, particularly those that are based on sum frequency generation (SFG), are in principle well suited for probing water near membranes, this method has been largely limited to probing lipid monolayers[1, 7-16] chemically asymmetric bilayers,[17-19] or the use of D$_2$O as opposed to H$_2$O.[20-21] Indeed, the use of SFG spectroscopy for probing fully hydrogenated lipid bilayers is now just emerging. Part of the reason for this relatively new application of vibrational SFG spectroscopy to probe chemically unmodified lipid bilayers is rooted in the symmetry-breaking requirement of the method,[22] which has limited its use largely to asymmetric bilayers consisting of a deuterated and a hydrogenated leaflet, or lipid monolayers, as stated above. SFG signals generated by asymmetric membranes (deuterated leaflet on one side and hydrogenated leaflet on the other side, or aliphatic lipid tail on one side and polar headgroup on the other) are strong enough to be detectable using low-repetition rate, low peak power laser systems most commonly used in the field. Two studies known to us also report SFG spectra of unlabeled symmetric lipid bilayers, demonstrating their low signal yields when compared to labeled bilayers.[23-24] Our recent work[25-27] has shown, in the C–H stretching region, that commercially available broadband optical parametric amplifier laser systems running at modest (kHz) repetition rates can overcome these limitations, with reasonably high signal-to-noise ratios obtained in just a few minutes of spectral acquisition time.

Here, we report how to apply this approach to probe the C–H stretches of the alkyl tails in fully hydrogenated single- and dual-component supported lipid bilayers (SLBs) along with the O–H stretching continuum of the H-bond network system in the electrical double layer above them. The approach probes lipid tail order and disorder while also informing on changes in the H-bond network strength that result from changes in the bulk ionic strength up 100 mM NaCl. Moreover, by varying the lipid bilayer composition from 100% zwitterionic lipid to an 8:2 mixture of zwitterionic and negatively charged lipids, we identify specific H-bond interactions between water molecules and the lipid headgroup choline moieties that manifest themselves in spectral intensity changes in the 3100 cm$^{-1}$ to 3200 cm$^{-1}$ range.

## II. Methods.

**A. Bilayer Preparation.** 1,2-dimyristoyl-*sn*-glycero-3-phosphocholine (DMPC) and 1,2-dimyristoyl-*sn*-glycero-3-phospho-(1-*rac*-glycerol) (DMPG) were purchased from Avanti Polar Lipids and used without further purification. Lipid bilayers from small unilamellar vesicles of pure DMPC, lipid mixtures containing 90 mol% DMPC and 10 mol% DMPG, and 80 mol% DMPC and 20 mol% DMPG were prepared by the vesicle fusion method, as described earlier,[25, 27-30] on 3 mm thick calcium fluoride windows (ISP Optics, CF-W-25-3). Prior to use, the calcium fluoride window was sonicated in HPLC-grade methanol (Fisher Scientific) for 30 min, rinsed with ultrapure water (18.2 $\Omega$·cm resistivity; Millipore), and dried with $N_2$. The window was then plasma cleaned (Harrick Plasma Cleaner, 18W) for 10 min.

Experiments were carried out at room temperature (21 ± 2 °C). All SLBs were formed at 0.01 M Tris buffer and 0.1 M NaCl in the presence of 0.005 M $CaCl_2 \cdot 2H_2O$ at pH 7.40 ± 0.03.[29] Following bilayer formation, SLBs were rinsed with Ca-free buffer to remove excess vesicles. The spectra were recorded at two different ionic strengths. Before the preparation of aqueous solutions, Millipore water was left overnight to equilibrate with atmospheric $CO_2$. The solution pH was measured for each salt concentration and the pH was adjusted to 7.4 with minimal NaOH and HCl before the solutions were flowed across the interface resulting in ionic strengths of ~10 μM and 0.1 M for the Millipore solution and NaCl solution, respectively.

**B. Vibrational Sum Frequency Generation Spectroscopy.** Details of our SFG approach and experimental setup for probing condensed matter interfaces in the C–H stretching region have been reported previously.[29-33] Here, we adapt this approach to extend our spectral range into the O–H stretching region, as described in detail in the Supporting Information (see part numbers of the optical elements in Supplementary Figure S1). Briefly, 90% of the output from a Ti:Sapphire amplifier laser system (Spectra Physics Solstice, 3 mJ/pulse, 795 nm pulses, 1 kHz repetition rate, 120 femtosecond pulse duration) pumps a travelling-wave optical parametric amplifier to generate a broadband tunable IR beam tuned to the C–H and O–H regions (2800-3600 cm$^{-1}$), while the remaining portion is sent down the visible upconverter beam line, where it is attenuated using a variable density filter and spectrally narrowed using an etalon. The IR and visible beams are focused to a ~30 μm beam waist at the interface, where they overlap at the CaF$_2$/water interface at 38° and 30° from the surface normal, respectively. The resultant SFG signal is dispersed on to a spectrograph (Acton SP-2558) and liquid nitrogen-cooled CCD camera. During the SFG experiments, the IR line was purged with dry house N$_2$ to avoid water absorption bands that appear in this stretching region. All SFG spectra were recorded in triplicates and normalized to the *ppp*-polarized SFG response obtained from a gold window. Further details regarding spectral acquisition and analysis procedures are provided in the Supporting Information (See Figure S2).

**C. FRAP Measurements.** Two-dimensional diffusion coefficients, which can serve as a metric for bilayer quality, were estimated using fluorescence recovery after photobleaching (FRAP). FRAP measurements and sample preparation were carried out in a manner consistent with our previous approach.[25] For these experiments, vesicles composed of DMPC or a 9:1 mixture of DMPC/DMPG lipids were doped with 0.1 mol% TopFluor PC (Avanti Polar Lipids, 810281). After forming the SLB as described in Section IIA, the cell was flushed with 20 mL of 0.1 M NaCl, 0.01 M Tris buffer (pH 7.4). In a second set of experiments, the flow cell was flushed with 20 mL of pH-adjusted Millipore water with no added salt. For SLBs formed from a 9:1 mixture of DMPC/DMPG lipids, we find diffusion coefficients on the order of 0.5 ± 0.2 μm$^2$/s (13 replicates over two samples) after rinsing with 0.1 M NaCl, 0.01 M Tris buffer, which is consistent with our previously reported two-dimensional diffusion coefficients[25] and indicates that a well-formed bilayer is produced from the abovementioned method.[34-36] Upon rinsing with pH-adjusted Millipore water with no added salt, we find that the diffusion coefficient for SLBs formed from 9:1 mixtures of DMPC/DMPG lipids are on the order of 0.03 ± 0.01 $\mu m^2$/s (6 replicates over 1 sample). The diffusion coefficient for SLBs formed from pure DMPC lipids on calcium fluoride is 0.4 ± 0.2 $\mu m^2$/s (6 replicates over two samples) after rinsing with 0.1 M NaCl, 0.01 M Tris buffer. After rinsing with pH adjusted Millipore water, we find a diffusion coefficient of 0.07 ± 0.02 $\mu m^2$/s (4 replicates over two samples). For SLBs formed from 8:2 mixtures of DMPC/DMPG lipids, the diffusion coefficient on calcium fluoride is 0.07 ± 0.04 $\mu m^2$/s (6 replicates over one sample) after rinsing with 0.1 M NaCl, 0.01 M Tris buffer. Representative traces, along with a detailed procedure used in these experiments, are provided in the Supporting Information (see Figure S3). These results indicate the bilayers transition between the gel and fluid phases, irrespective of the nature of the underlying substrates ($CaF_2$ vs fused silica).

**D. Computational Methods.** Molecular dynamics (MD) simulations for investigating the structure of the H-bond network near the lipid-water interface were performed using the CHARMM-GUI[37] input generator to set up the DMPC and 9:1 DMPC/DMPG systems. Each system contains a 10 $\times$ 10 $nm^2$ lipid bilayer. For pure DMPC, systems were set up with 0.15 M NaCl or no salt added, both with a hydration level (i.e., water:lipid ratio) of 53. The 9:1 DMPC/DMPG system was set up with 0.15 M NaCl and a hydration level of 65. We performed equilibration and production runs with CHARMM-GUI generated input files using the NAMD[38] package. The CHARMM36[39-40] force field was applied for the lipid, water, and ions. The Particle-Mesh-Ewald[41] (PME) method was used for the electrostatic interactions with a real-space cutoff of 1.2 nm. Force switching with a cutoff of 1.2 nm was applied to the van der Waals interactions. The PME grid size was set to 108, 108, and 100 for the X, Y, and Z dimensions in the DMPC simulations, and to 108, 108, and 120 for the 9:1 DMPC/DMPG simulations. RATTLE[42] was applied to constrain all bonds involving hydrogen atoms in length. Langevin dynamics were applied for constant pressure and temperature control. A Nose-Hoover Langevin piston[43-44] was applied with constant ratio on the X-Y plane and a target pressure of 1 atm. The target temperature was set to be 303.15 K with a damping coefficient of 1.0 ps$^{-1}$. For the DMPC systems, the production run lasted for 30 ns with a 2 fs time step; for the 9:1 DMPC/DMPG systems, the production was run for 70 ns. Any unspecified details, including the equilibration process before production runs, are consistent with the standard CHARMM-GUI protocol.

### III. Results and Discussion.

**A. Single-Component Zwitterionic Supported Lipid Bilayers.** Figure 1A shows the *ssp*-polarized SFG response from the pure DMPC bilayer without added salt. At this low ionic strength (~10 μM), we find clear spectral signatures from the C–H oscillators of the alkyl tails, [25, 27, 30] as well as broad contributions from the O–H stretches of the water molecules. The non-zero signals are due to the fact that the molecular environment above and below the bilayer is not fully symmetric, as would be expected for a suspended bilayer. Instead, symmetry breaking occurs due to the presence of the aqueous phase on one side and the solid support on the other.

The frequencies corresponding to the signal peaks in the C–H stretching region shown in Figure 1A are comparable to the ones we observe for supported lipid bilayers formed on fused silica substrates (see Supporting Information Figure S4)[25, 27, 30] The two broad features in the O–H stretching continuum located at ~3200 cm$^{-1}$ and ~3400 cm$^{-1}$ are associated with bandwidths (full width at half maximum) of about 200 cm$^{-1}$. The peak positions are within 50 cm$^{-1}$ of what has been reported for water spectra obtained from symmetric bilayers prepared from negatively charged lipids on $CaF_2$.[24] The difference is attributed to the fact that our current experiments use bilayers formed from purely zwitterionic lipids.

Replacing the $H_2O$ phase with $D_2O$ while maintaining low ionic strength, shown in Supporting Information Figure S5, leads to the C–H oscillators retaining their frequencies while the O–H stretching continuum is entirely absent. This experiment indicates that 1) there are no exogenous photon sources contributing to the SFG response from the bilayer under water ($H_2O$), and 2) that $H_2O$ that may be possibly trapped between the bilayer and the substrate is readily exchanged or associated with too little SFG intensity to be detectable by our method. Control experiments assessing the possible role that $CaF_2$ dissolution could have on the spectra[45-46] (see Supporting Information Figure S6) show that the presence of the bilayer eliminates any flow-dependent changes in the SFG signal intensity produced by the interfacial water molecules.

The O–H stretching continuum can be viewed as a display of the various O···O distances sampled in the water network probed by the SFG spectrometer. As shown, for instance, by Lawrence and Skinner,[47] frequencies around 3200 $cm^{-1}$ correspond to O–H stretches associated with water molecules in tighter H-bond networks, where H–O distances are as short as 1.6 Å or less. Towards 3400 $cm^{-1}$, the spectrum samples water molecules in a considerably looser H-bond network, having H–O distances as long as 2.1 Å or so. Towards 3550 $cm^{-1}$, H–O distances can be as long as 2.4 Å or more. At the very end of the spectrum, near 3700 $cm^{-1}$, would be the O–H stretch of non-H-bonded water molecules, those that "straddle the interface".[48] Such signals are not identified within our signal-to-noise ratio, even though they have been reported to be present in Langmuir monolayers prepared from DPPC lipids.[49]

Figure 1B shows the SFG spectrum from the supported lipid bilayer in comparison with that of two other aqueous $CaF_2$ interfaces, namely that of bare $CaF_2$ in contact with ~10 μM ionic strength water adjusted to pH 7.4, as well as bare CaF$_2$ in contact with water vapor in He flow adjusted to 80% relative humidity (see Supporting Information Figure S7). The SFG response from the bare CaF$_2$/water interface is in reasonable agreement with published data.[46, 50-51] We find that the peak positions from the bilayer/water interface is blue-shifted by around 25 cm$^{-1}$ when compared to those obtained from the bare CaF$_2$/water interface. Additionally, the SFG spectrum from the CaF$_2$/water vapor interface exhibits a blue-shifted SFG spectrum when compared to the bilayer/water or CaF$_2$/water interfaces, consistent with the expectation that its hydrogen-bonding environment is looser than in the case of bulk water in contact with the solids.[52-53]

Upon increasing the ionic strength in the bulk aqueous phase, the sodium and chloride ions can modify the H-bond network of water molecules in the bulk in ways that are the subject of much past and ongoing scientific attention and discussion.[54-55] NaCl, whose anion and cation fall right in the middle of the familiar Hofmeister series, are not necessarily expected to modify the H-bond network over lipid bilayers at the relatively modest concentrations (0.1 M) employed here. Moreover, ion-specific interactions with the lipids used in our work are unlikely to be strong under the conditions of our experiments. Indeed, Figure 2 shows that the spectral changes we observe in response to changes in the ionic strength are largely uniform over the entire frequency region probed in our experiment (1000 cm$^{-1}$). Between 3000 cm$^{-1}$ and 3600 cm$^{-1}$, the ratio of the SFG spectral intensities at low (~10 μM) and high (0.1 M) ionic strength is computed to vary only slightly, from 1.7 at 3000 cm$^{-1}$ to 2.3 at 3600 cm$^{-1}$ and back to 2.0 at 3700 cm$^{-1}$ (average of 2.1 ± 0.2 over all frequencies). We find this slight frequency dependence of the SFG intensity ratio to be indicative of a minor influence that the relatively modest salt concentrations used here even under what we term "high salt" have on the various contributors to the H-bond network.

This interpretation is borne out in molecular dynamics simulations as well, which are described next.

To further explore the molecular details near the bilayer/water interface, we performed MD simulations for a DMPC lipid bilayer with and without 0.15 M NaCl salt. We focus here on the analysis of the interfacial water structure, specifically the orientation of interfacial water molecules and the O⋯O distance of neighboring water molecules. The water orientation is characterized with the dipole angle, θ, which is defined as the angle between the dipole vector of water and the membrane normal pointing towards the bulk. The distribution of water orientations is analyzed as a function of distance from the membrane-water interface, *i.e.*, we plot the two-dimensional distribution[56] (Figure 3):

$$\omega(\theta, z) = \frac{\langle \delta(\theta - \theta(t)) \delta(z - z(t)) \rangle}{\rho(z) sin\theta} \qquad (1),$$

in which $\theta$ is the dipole angle defined above, $z$ is the normal distance of the water oxygen from the bilayer center, $\rho(z)$ is the number density of water, and $sin\theta$ is the angular Jacobian factor. The distribution shown in Figure 3 is normalized to that of the bulk value.

According to the mass density distribution (see Figure S8), the lipid-water interface is identified at $z \sim 20$ Å. As shown in Figure 3 (left column), in all cases studied, the water orientation distribution shifts towards smaller dipole angles near the interface, while the opposite shift is observed for the small amount of water molecules that penetrate below the lipid/water interface to interact with the lipid glycerol groups (for a snapshot, see Figure S9). The distribution approaches the bulk value at ~8-10 Å away from the lipid-water interface. Nevertheless, the distribution of water orientation remains broad even at the interface, which is likely due to the dynamic nature of the lipid headgroup (see Figure S10). As a result, no statistically significant difference is observed between the two DMPC cases studied, suggesting that the impact of salt on the water orientation at the interface is subtle compared to the effect of thermal fluctuations. Regarding the distributions of the nearest O···O distances among water molecules, which reports on the hydrogen bonding strength, our results in Figure 3 (right column) suggest again that the impact of salt is small, supporting the observation from SFG analysis.

The SFG signal intensity reductions observed across the frequency range investigated here as the salt concentration is raised from 10 μM to 0.1 M concentration levels are consistent with absorptive-dispersive mixing between $\chi^{(2)}$ and $\chi^{(3)}$ contributions to the SFG signal generation process from charged interfaces according to[57-60]

$$\chi^{(2)}_{total} = \chi^{(2)}_{NR} + \chi^{(2)}_{surf} + \frac{\kappa}{\sqrt{\kappa^2 + (\Delta k_z)^2}} e^{i\,arctan\left(\Delta k_z/\kappa\right)} \Phi(0)\chi^{(3)} \quad (2)$$

Here, the first two terms are the non-resonant and resonant 2nd-order susceptibility and the 3rd term is given by the inverse Debye screening length, $\kappa$, the inverse of the coherence length of the SFG process, $\Delta k_z$, and the interfacial potential, $\Phi(0)$, multiplied by the 3nd-order susceptibility. The $\chi^{(3)}$ phase angle $\varphi = arctan\left(\Delta k_z/\kappa\right)$ can be estimated from Gouy-Chapman theory: at the low (*resp*. high) salt concentration investigated here, $\kappa$ is 1 x $10^7$ (*resp*. 1 x $10^9$) m$^{-1}$, while our experimental geometry leads to a $\Delta k_z$ of 2.4 x $10^7$ m$^{-1}$, which is invariant with salt concentration. The resulting phase angle is shown in Figure 4A.

In the absence of phase-resolved measurements, which are proving to be considerably challenging at buried liquid-solid interfaces such as the ones studied here, it is difficult to quantitatively examine the interfacial potential, even if one uses the 3rd order ($\chi^{(3)}_{bulk}$) term recently reported by Wen et al.[57] that should be quite universally applicable for aqueous interfaces. Moreover, it is perhaps not possible to prepare, in an experiment, a truly "zero potential" reference state: even the fully protonated reference state of a carboxylic acid monolayer, commonly used as a reference state in surface potential measurements,[57, 61-62] is subject to dipolar potentials. In the absence of 1) phase resolved data and 2) a true zero potential – and/or zero charge density – reference state, quantitative knowledge of the interfacial potential at two different solution or bilayer conditions from which a difference in surface potential, i.e. $\Delta\Phi$, can be calculated is difficult to obtain, though methods to acquire this knowledge remain a topic of keen interest to us that we will discuss in forthcoming work.

For now, we offer the following method for estimating surface potential changes from the second-order spectral lineshapes (in the OH stretching region): an examination of Equation 2 reveals that even if, as suggested by the MD simulations discussed above, the H-bond network close to the interface remains invariant or nearly invariant (implying a constant $\chi^{(2)}_{surf}$) upon changes in ionic strength, changes in the SFG signal intensity can still arise from the potential-dependent $\chi^{(3)}$ term. These changes take the form of a complex multiple of the $\chi^{(3)}_{bulk}$ term, which is given mainly by the $3^{rd}$ order optical properties of bulk water. Unfortunately, given the difficulties discussed above, our lack of phase-resolved measurements and our lack of access to a reference state of true $\Phi(0)=0$, precludes us from comprehensively accounting for the phase-angle dependent $\chi^{(2)}/\chi^{(3)}$ mixing, and thus quantitatively determining the interfacial potential from the SFG spectra reported here. Yet, surprisingly good qualitative agreement is obtained between the difference of the measured intensity spectra for the low and high salt conditions from Figure 2 and the calculated $\chi^{(3)}_{bulk}$ intensity spectrum derived from the real and imaginary data reported by Wen *et al.* (see Figure 4). This agreement supports our conclusion that the spectral changes are not indicative of large changes in the H-bonded network of water molecules but rather result from the $\chi^{(3)}$-potential dependent term. Moreover, as shown Supplementary Information equations S1-S5, under conditions where the SFG responses are dominated by the $\chi^{(3)}$ term, *i.e.*

$\chi^{(3)} \Phi >> \chi^{(2)}$, an estimate of the difference in surface potential, $\Delta\Phi$, can be readily provided if the magnitude of the SFG intensity difference, $\Delta I_{SFG}$, observed for conditions of varying ionic strength, bulk solution pH, analyte concentration, or surface composition, is known (see Supporting Information Eqn S5).

**B. Dual-Component Supported Lipid Bilayers Formed from Zwitterionic and Negatively Charged Lipids.** Motivated by recent reports that the major contribution in the 3000 cm$^{-1}$ to 3200 cm$^{-1}$ frequency region originates from polarized water molecules that bridge phosphate and choline in the zwitterionic lipid headgroup (*n. b.*: that work focused on lipid monolayer/water interfaces as opposed to lipid bilayer/water interfaces, which are probed in the present study),[2] we proceeded to add negatively charged lipids to the zwitterionic system studied. Mixing in negatively charged lipids, such as DMPG, is then expected to reduce the population of polarized water molecules that interact specifically with the zwitterionic PC headgroup.

Figure 5 shows that this response is indeed observed. At 0.1 M NaCl, the three systems we surveyed (100% zwitterionic DMPC, 9:1 DMPC/DMPG, and 8:2 DMPC/DMPG) showed no significant changes in the 3400 cm$^{-1}$ frequency region. Yet, as the percentage of negatively charged lipids increases, the SFG spectral intensity in the 3200 cm$^{-1}$ region decreases, indicating the theoretical result obtained for lipid monolayer/water interfaces may also hold for lipid bilayer/water interfaces. Triplicate measurements are shown in the Supporting Information (see Figure S11).

Results from our MD simulation for 9:1 DMPC/DMPG with 0.15 M NaCl (Figure 3, bottom row) reveal similar trends when compared to the pure DMPC case, suggesting that the impact of a small amount (10%) of anionic lipids on the structure and orientation of water at the interface is minor, in the background of thermal fluctuations. Yet, computing the number of water molecules adjacent to lipid phosphate, choline, and those close to both phosphate and choline (see Figure S12 for the relevant radial distribution functions), in a manner consistent with the analysis by Morita and coworkers,[2] we find that mixing in DMPG lipids leads to a small, albeit statistically significant, decrease in the number of water molecule adjacent to both the lipid phosphate and choline moieties per area, as shown in Table 1. These computational results support the observations that the SFG signal intensities seen in the experimental spectra between 3100 cm$^{-1}$ to 3200 cm$^{-1}$ are due to local water molecules that specifically interact with the phosphate and choline moieties of the DMPC lipids.[2] As shown in Figure S13, these water molecules are also subject to a fairly broad molecular orientation distributions (with the second moment of the dipole angle θ in the range of 34-38°) due to thermal fluctuations at the lipid/water interface.

**IV. Conclusion.** In conclusion, we have reported vibrational sum frequency generation spectra in which the C–H stretches of lipid alkyl tails in fully hydrogenated single- and dual-component supported lipid bilayers are detected along with the O–H stretching continuum of the hydrogen-bond network in the electrical double layer above the bilayers. Aided by atomistic simulations, we find that the hydrogen bond network over the purely zwitterionic bilayers is largely invariant with salt concentration between sub-micromolar and 100s of millimolar concentrations. The structure of the lipid tails are largely invariant with salt concentration as well, as indicated by a lack of relative spectral changes in the SFG responses observed in the C–H stretching region as salt concentration is varied. As the salt concentration is increased from ~10 μM to 0.1 M, the SFG intensities in the O–H stretching region decrease by a factor of 2. This observed salt concentration-dependent change in the SFG signal intensity is consistent with significant absorptive-dispersive mixing between $\chi^{(2)}$ and $\chi^{(3)}$ contributions to the SFG signal generation process from charged interfaces. Surprisingly good qualitative agreement is obtained between the difference of the measured intensity spectra for the low and high salt conditions from Figure 2 and the calculated $\chi_{bulk}^{(3)}$ intensity spectrum derived from the real and imaginary data reported by Wen *et al.* (Figure 4). This agreement supports our conclusion that the spectral changes are not indicative of large changes in the H-bonded network of water molecules but rather result from the $\chi^{(3)}$-potential dependent term. Moreover, our analysis provides a method for estimating the difference in surface potential, $\Delta\Phi$, from the magnitude of the SFG intensity difference, $\Delta I_{SFG}$, observed for conditions of varying ionic strength, bulk solution pH, analyte concentration, or surface composition, is known (see Supporting Information Eqn S5).

Specific interactions between water molecules and lipid headgroups are observed as well: Replacement of PC lipids with negatively charged PG lipids coincides with SFG signal intensity reductions in the 3100 cm$^{-1}$ to 3200 cm$^{-1}$ frequency region. Our atomistic simulations show that this outcome is consistent with a small, albeit statistically significant, decrease in the number of water molecules adjacent to both the lipid phosphate and choline moieties per unit area, supporting the SFG observations. This result further supports recent molecular dynamics simulations indicating that the major contribution in the 3000 cm$^{-1}$ to 3200 cm$^{-1}$ frequency region originates from polarized water molecules that bridge phosphate and choline in the zwitterionic lipid headgroup.[2] Ultimately, the ability to probe H-bond networks over lipid bilayers holds the promise of opening paths for understanding, controlling, and predicting specific and non-specific interactions membranes with solutes such as ions and small molecules, peptides, polycations, proteins, and coated and uncoated nanomaterials.

**Acknowledgements.** This work is supported by the National Science Foundation under the Center for Sustainable Nanotechnology, Grant No.CHE-1503408. PEO and ACM gratefully acknowledge support from the U.S. National Science Foundation Graduate Research Fellowship Program. PEO is a Northwestern University Presidential Fellow. FMG gratefully acknowledges support from a Friedrich Wilhelm Bessel Prize from the Alexander von Humboldt Foundation.

# References.

**Figure Captions.**

**Figure 1. (A)** *ssp*-Polarized SFG spectrum of an SLB made from pure DMPC lipids in contact with low-ionic strength water adjusted to pH 7.4. The data below 3000 cm$^{-1}$ have not been binned to preserve the C–H stretching region, while they were binned over nine points in x and y between 3000 cm$^{-1}$ and 3800 cm$^{-1}$. **(B)** Comparison of *ssp*-polarized SFG spectra from the CaF$_2$/wet air interface (gray), the CaF$_2$/water interface (blue), and a CaF$_2$-supported SLB prepared from pure DMPC lipids (green). Lines represent spectra binned by over nine points in x and y between 2800 cm$^{-1}$ and 3800 cm$^{-1}$.

**Figure 2.** *ssp*-Polarized SFG spectrum of an SLB formed from pure DMPC lipids in Millipore water with no added salt (dashed lines) and with 100 mM NaCl (solid lines) at 21 ºC and pH 7.4. The lines represent the data that have been binned by over nine points in x and y between 3000 cm$^{-1}$ and 3600 cm$^{-1}$.

**Figure 3.** Left column: Two-dimensional orientational distribution of water ($\omega(\theta,z)$) near the lipid bilayer/water interface from atomistic MD simulations. Right column: Two-dimensional distribution of nearest O$^{\cdots}$O distance of water near the lipid bilayer/water interface. Lipid composition and ionic strength are shown in the legend. The z axis represents the membrane normal, and z = 0 is located at the center of bilayer. Each analysis is averaged over 1000 frames. Results are for DMPC with no salt (top), DMPC with 0.15 M NaCl (center), and 9:1 DMPC/DMPG with 0.15 M NaCl (bottom).

**Figure 4**. **(A)** Variation of the $\chi^{(3)}$ phase angle $\varphi = arctan\left(\frac{\Delta k_z}{\kappa}\right)$ with ionic strength under the conditions of our experiments. **(B)** Potential dependent $\chi^{(3)}$ effect: comparison between the difference between the low and high salt conditions from Figure 2 (green) and a calculated intensity spectrum derived from the $\chi^{(3)}$ spectra reported by Wen *et al.*[57] (blue). Even without comprehensively accounting for the absorptive-dispersive mixing controlled by the $\chi^{(3)}$ phase angle, $\varphi$, as per equation (2), which is not yet accessible through phase-resolved measurements at the solid/liquid interface, qualitative agreement is demonstrated.

**Figure 5.** *ssp*-Polarized SFG spectrum of an SLB formed from pure DMPC (dark green), a 9:1 mixture of DMPC/DMPG (light green) and an 8:2 mixture of DMPC/DMPG (blue) lipids in 100 mM NaCl (solid lines) at 22 ºC and pH 7.4. The circle markers represent the raw data while the darker lines represent the data that have been binned by over nine points in x and y between 3000 cm$^{-1}$ and 3600 cm$^{-1}$.

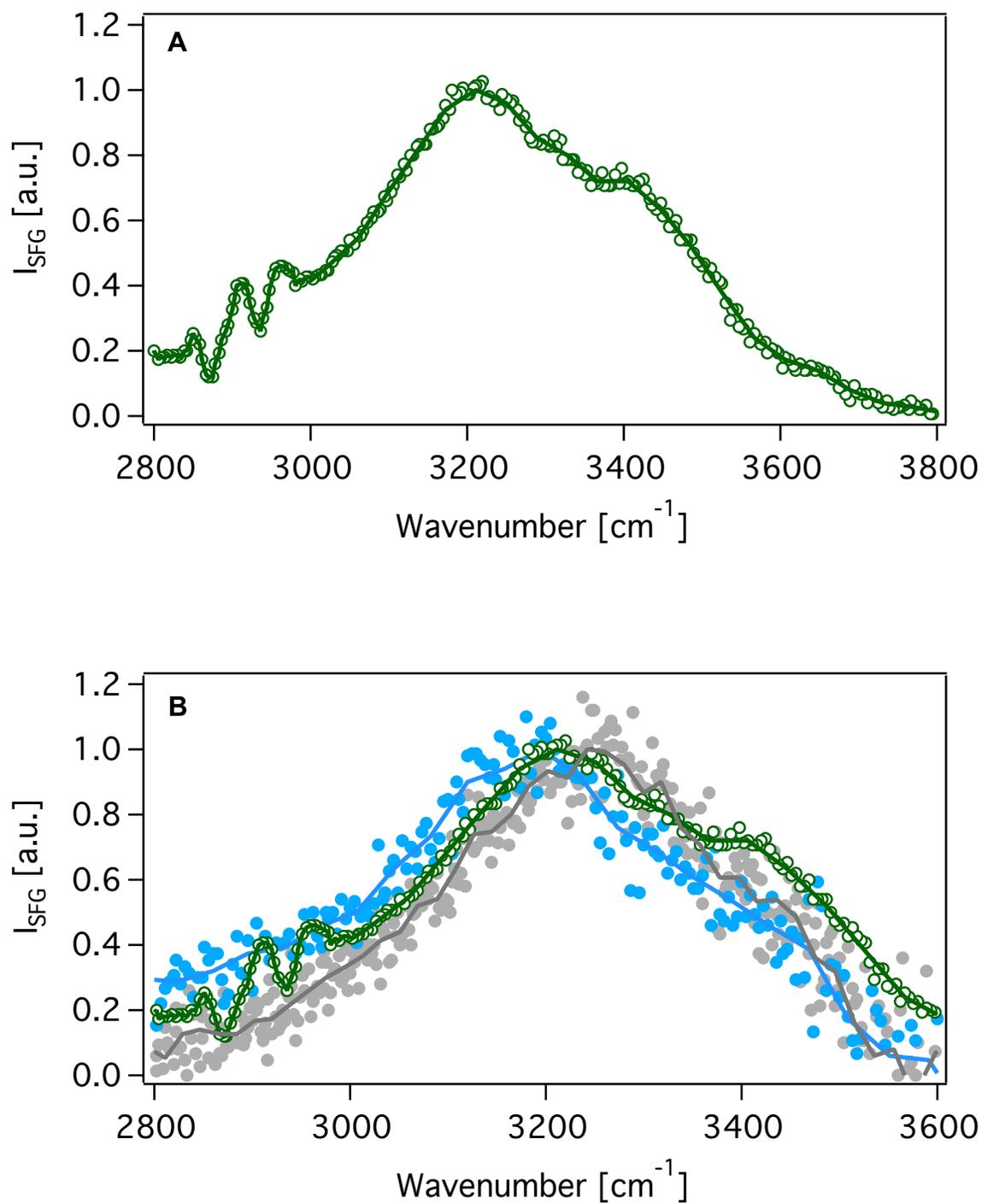

**Figure 1.**

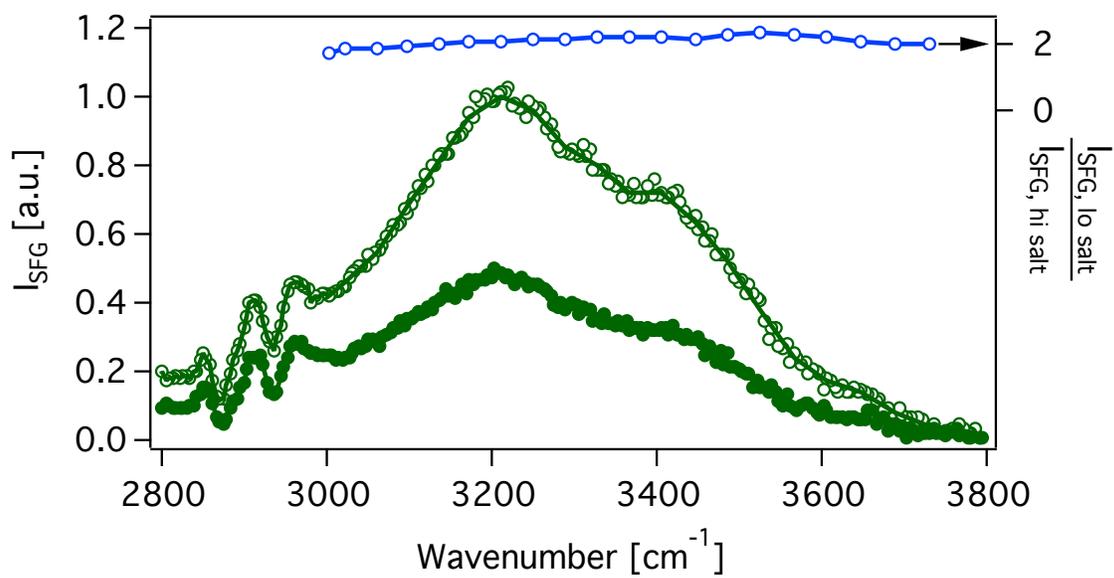

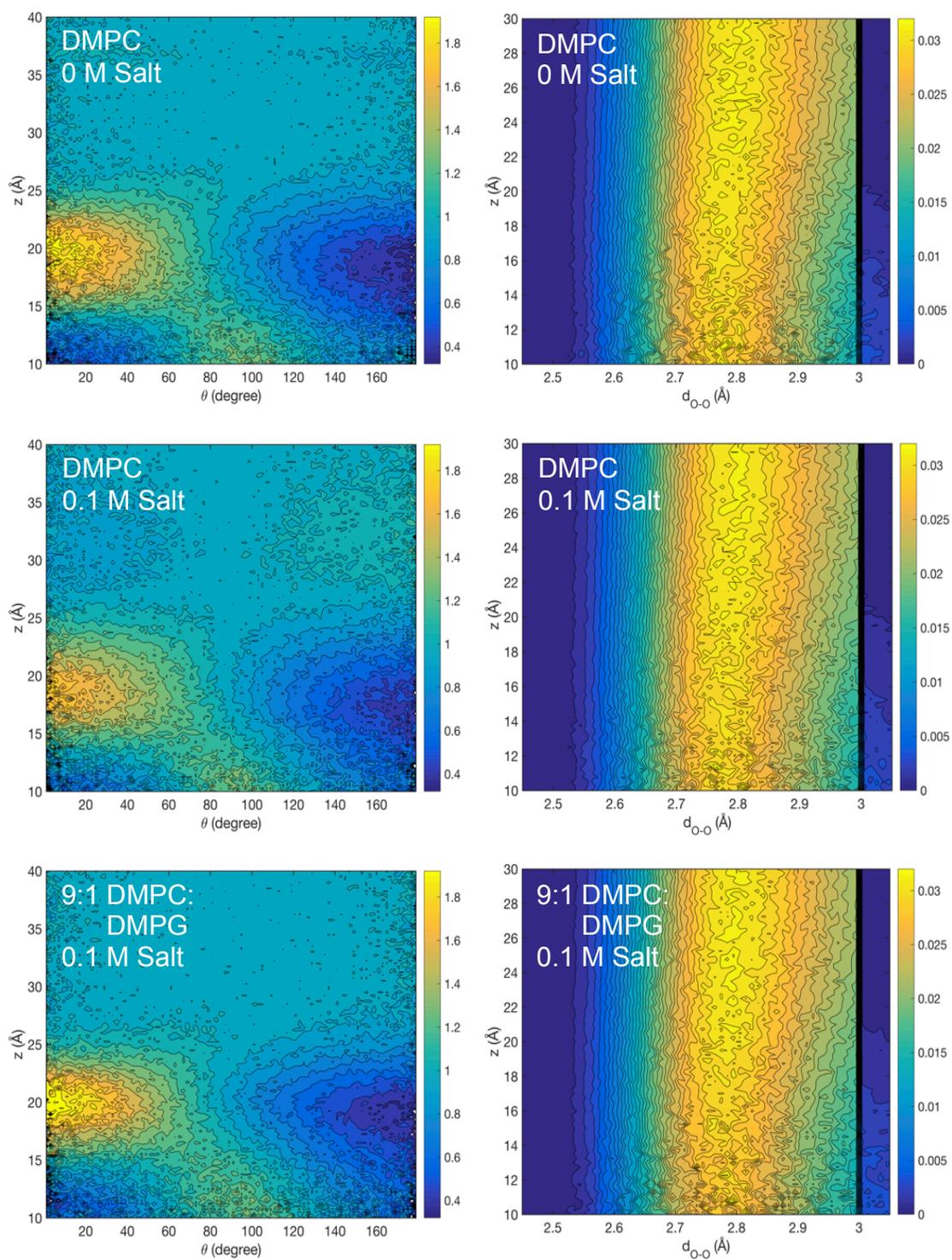

**Figure 3.**

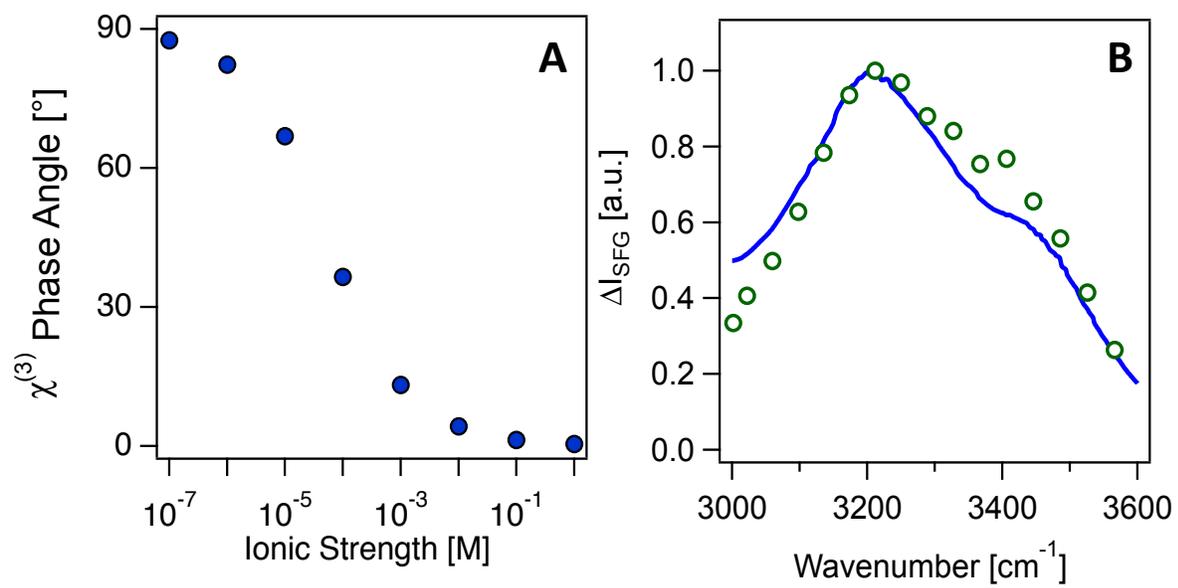

**Figure 4.**

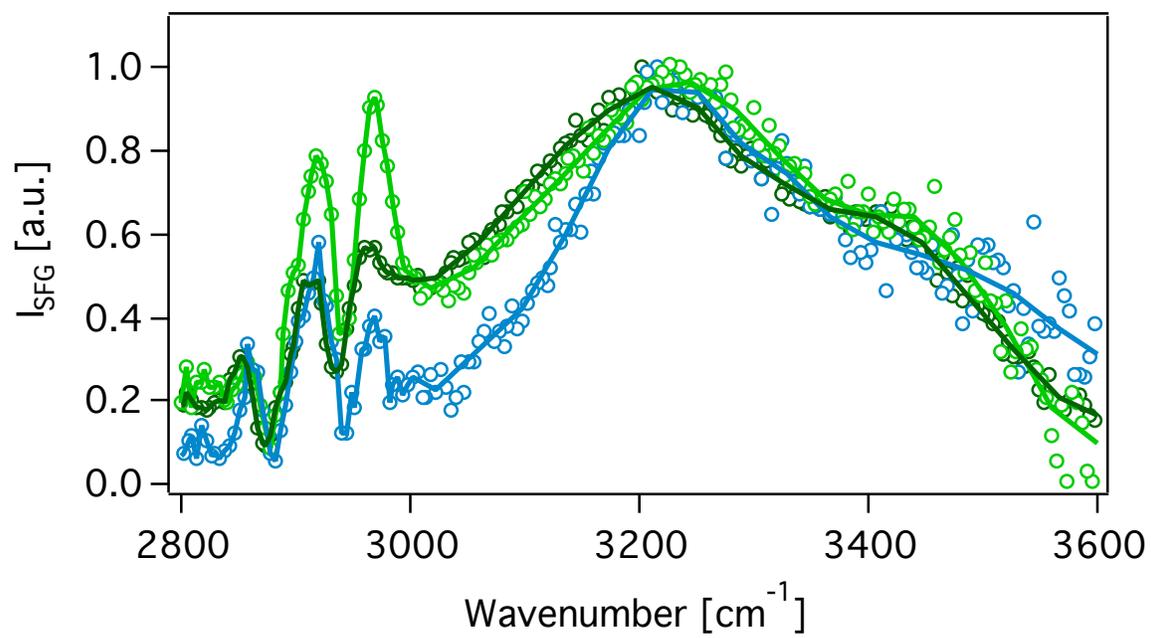

**Figure 5.**

**Table 1.** Number of water molecules (per surface area) close to the lipid phosphate (labeled with "P"), close to choline (labeled with "N"), and close to both phosphate and choline groups (labeled with "NP") from atomistic MD simulations.[a]

| Lipid composition | DMPC | | 9:1 DMPC/DMPG |
|---|---|---|---|
| [NaCl] | 0 M | 0.15 M | 0.15 M |
| # $H_2O@P$ (nm$^{-2}$) | $20.4 \pm 0.3$ | $20.5 \pm 0.3$ | $20.4 \pm 0.2$ |
| # $H_2O@N$ (nm$^{-2}$) | $19.4 \pm 0.2$ | $19.9 \pm 0.3$ | $18.6 \pm 0.2$ |
| # $H_2O@N\&P$ (nm$^{-2}$) | $15.9 \pm 0.3$ | $16.1 \pm 0.2$ | $15.2 \pm 0.2$ |

[a]Following the work of Morita and co-workers,[2] the cutoff distance for water near phosphate is determined by the second minimum of the O–P radial distribution function; the cutoff distance for water near choline is determined by the first minimum of the O–N radial distribution function (see Figure S9).

**TOC Graphic**

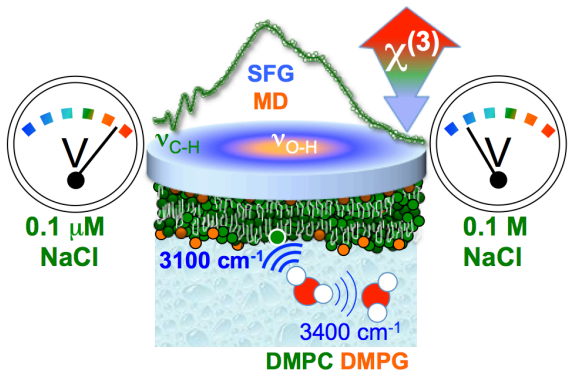